# Plasmonics of coupled graphene micro-structures

Hugen Yan[1,]*, Fengnian Xia[1], Zhiqiang Li[2], and Phaedon Avouris[1],*

[1]IBM T. J. Watson Research Center, Yorktown Heights, NY 10598, USA
[2]National High Magnetic Field Laboratory, Tallahassee, Florida 32310, USA
*To whom correspondence should be addressed
Emails: hyan@us.ibm.com (H. Y.), avouris@us.ibm.com (P. A.)

Abstract

The optical response of graphene micro-structures, such as micro-ribbons and disks, is dominated by the localized plasmon resonance in the far infrared (IR) spectral range. An ensemble of such structures is usually involved and the effect of the coupling between the individual structures is expected to play an important role. In this paper, the plasmonic coupling of graphene microstructures in different configurations is investigated. While a relatively weak coupling between graphene disks on the same plane is observed, the coupling between vertically stacked graphene disks is strong and a drastic increase of the resonance frequency is demonstrated. The plasmons in a more complex structure can be treated as the hybridization of plasmons from more elementary structures. As an example, the plasmon resonances of graphene micro-rings are presented, in conjunction with their response in a magnetic field. Finally, the coupling of the plasmon and the surface polar phonons of $SiO_2$ substrate is demonstrated by the observation of a new hybrid resonance peak around 500cm$^{-1}$.

## 1. Introduction

Graphene has very remarkable electrical[1], mechanical[2] as well as optical properties[3-6]. As a conductive atomic film, graphene has been predicted[7-9] and demonstrated to support surface plasmons in terahertz[10, 11] and infrared frequencies[12]. Compared to plasmons in noble metals[13, 14], which are widely studied and used, plasmons in graphene have two major advantages[7, 15]. First, the carrier density of graphene can be tuned by electrostatic or chemical doping, which provides the highly desired tunability to the plasmonic materials[10-12]. Second, due to the high mobility of carriers, plasmons in graphene can have long lifetime and low Ohmic loss. For instance, a lifetime of the order of ~100 femto-seconds has been demonstrated in graphene[11, 16], while in noble metals, the lifetime of the plasmon is in the sub-10 femto-second regime[13]. In addition, other advantages such as extreme light confinement, fabrication compatibility with current technologies and the availability of wafer-scale graphene make graphene a very promising plasmonic material which can complement existing plasmonic materials in a broad electromagnetic spectrum range.

The plasmonic coupling of artificial structures forms the foundation for optical metamaterials[17]. The coupling effect on the plasmon resonance in metallic nano-particle dimers[18], chains[19] and 2-dimensional arrays[20, 21] has been extensively studied. The plasmonic interactions in more complex metal structures, such as rings[22], concentric and non-concentric ring-disk[23] and three-dimensional structures[24] are still of great current interest. In this context, plasmon hybridization[25], Fano resonances[26], electromagnetically induced transparency (EIT)[27], radiative engineering of plasmon lifetime[21] and many other interesting phenomena are observed[14]. These studies enable plasmon applications in fields such as chemical and biological sensing[28], wave guiding[29], plasmon rulers[30] and field enhanced optical spectroscopy[31]. Despite its great importance, the plasmonic coupling in graphene microstructures has yet to be explored experimentally.

In this paper, we concentrate on the plasmonic interactions of graphene micro-structures. We will first discuss the interactions of graphene disks on the same plane (Section 2), followed by that of vertically stacked disks in the strong coupling limit (Section 3), where a fundamental difference between graphene plasmons involving massless Fermions and plasmons of charge carriers with mass is observed. Those two sections review and expand the discussion of previously published results[11]. Then plasmon hybridization[25] in graphene rings will be presented, together with its behavior in a magnetic field (Section 4). Finally, the signature of plasmon coupling to the substrate polar phonons is shown, supporting the notion that graphene is ultrasensitive to the environment (Section 5). Our study paves the way for applications of graphene in sensing, infrared photo-detection[32], light modulation[33], and terahertz metamaterials[10, 17].

**2. In-plane plasmonic coupling of graphene micro-disks**

Due to the momentum mismatch[9], light can not directly excite plasmons in graphene. However, the localized plasmon in micro-structured graphene, such as graphene disks (Fig. 1a), can be directly excited. Without considering the coupling of a disk to the adjacent disks, the optical conductivity of a disk array has a simple damped oscillator form[34],

$$\sigma(\omega) = i\frac{fD}{\pi}\frac{\omega}{(\omega^2 - \omega_p^2) + i\Gamma_p\omega} \qquad (1)$$

where $\omega$ is the frequency, $f$ is the filling factor (graphene area over total area), $D$ is the Drude weight[6, 35], and $\Gamma_p$ is the plasmon resonance width. The resonance frequency is[34, 36]:

$$\omega_p = \sqrt{\frac{3D}{8\varepsilon_m \varepsilon_0 d}} \quad (2)$$

where $\varepsilon_m$ is the media dielectric constant, $\varepsilon_0$ is the vacuum permittivity and $d$ is the diameter of the graphene disk. These formulae can be derived from the Drude conductivity of graphene[6, 37] and the polarizability of an ablate spheroid in the quasi-static limit[34, 36]. Neglecting multi-reflection effects from the substrate surfaces, the optical conductivity is related to the extinction spectrum of the sample through[38]:

$$1 - T/T_s = 1 - \frac{1}{\left|1 + Z_0 \sigma(\omega)/(1+n_s)\right|^2} \quad (3)$$

where $Z_0$ is the vacuum impedance, $n_s$ is the refractive index of the substrate, $T$ and $T_s$ are the transmissions through the sample and bare substrate respectively, as illustrated in Fig. 1b. When the extinction is small, it is directly proportional to the real part of the optical conductivity.

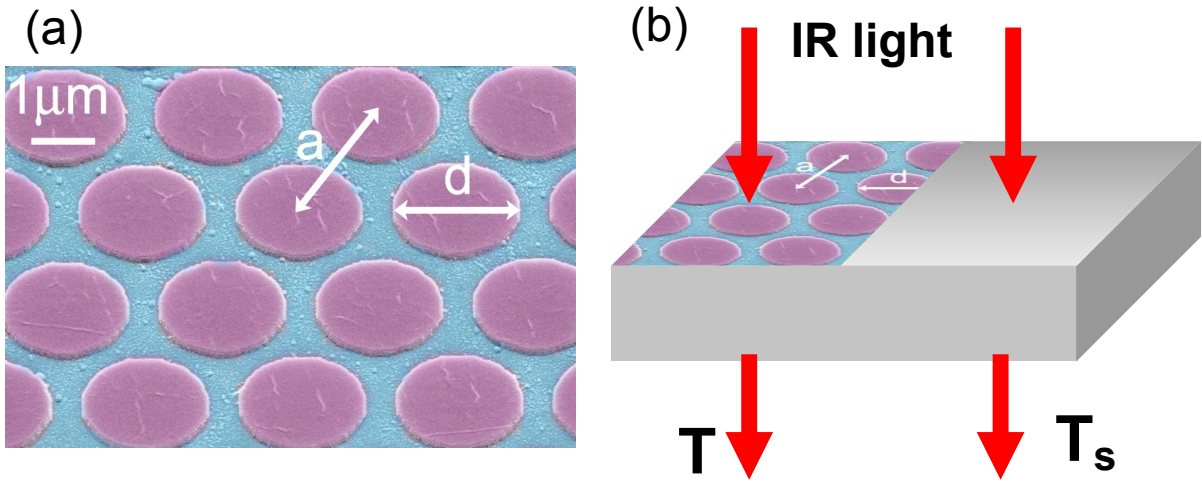

**Figure 1**. Samples and measurement configuration. (a) A scanning electron micrograph (SEM) of a typical graphene disk array. Lattice constant *a* and disk diameter *d* are indicated. (b) Transmission measurement configuration. The transmission spectrum is normalized by the transmission through the bear substrate ($T_s$).

In the experiments, chemical vapor deposition[39] grown graphene is transferred to $SiO_2$/Si or other substrates. Subsequently, e-beam lithography and dry etching are employed to define graphene disks or other geometric shapes, such as rings. Fig. 1a shows a typical SEM image of a disk array arranged in a triangular lattice. The transmission measurements are performed using a Fourier transform infrared spectrometer equipped with a silicon bolometer. The spectral range reported in this paper is in the far-IR/terahertz regime ($40cm^{-1}$-$550cm^{-1}$). Fig. 1b shows the measurement configuration.

Equations (1) and (3) can describe the lineshape of the extinction spectra of disk arrays quite well. The spectra in Fig. 2b and Fig. 3b can be fitted with those equations, with resonance frequency, linewidth and amplitude as fitting parameters. The lineshape agrees well with the model. However, the frequency of the resonance is not fully captured by equation (2), since the interaction between disks is not considered. Disk arrays with the same disk diameter but different lattice constants can have slightly different resonance frequencies. Indeed, this is what we observe. Fig. 2b shows the extinction spectra for three disk arrays with the same disk diameter (*d*=0.8μm) but three different lattice constants. With increasing lattice constant, the resonance frequency increases, and in the meantime, the amplitude decreases due to the smaller filling factor *f*. The result indicates that the coupling between disks in a triangular array softens the resonance frequency. This is not surprising since the dipolar oscillation, which is responsible for the plasmon resonance, can be screened by the nearby dipoles of adjacent disks. To view this in a simple picture, as shown in Fig. 2a, when two oscillating dipoles are brought together, the dipole-dipole interaction will weaken the restoring force of the oscillating charges in each disk, which results in lower oscillation frequency. Of course, exact quantitative description of the dipole-dipole interaction[40] is more complicated in our case, because many dipoles are involved in the array. Based on the plasmon interaction, the resonance frequency can be a measure of the disk distance. The concept of "plasmon ruler"[18, 30, 41, 42] is based on this and it is widely explored for metallic particle plasmons[14].

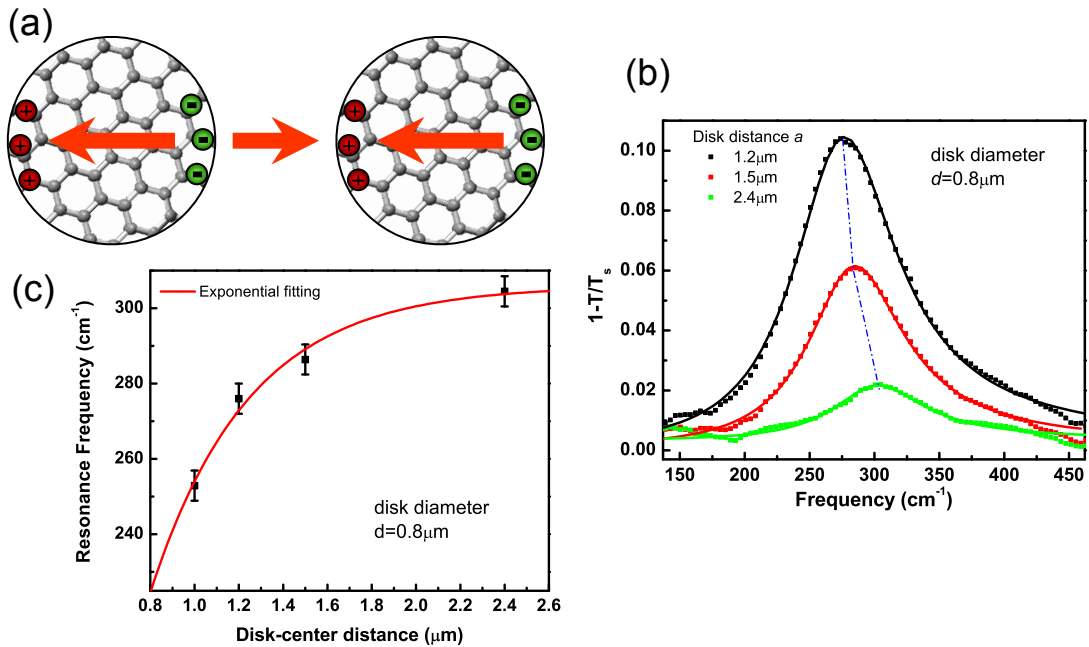

**Figure 2**. Plasmon interaction on the same plane. (a) An illustration of the plasmon coupling for two graphene disks on the same plane. In this specific case, the restoring force is weakened. (b) Extinction spectra for 3 arrays of graphene disks with different lattice constants $a$ but the same disk diameter $d$. The solid curves are fits using Equations (1) and (3). The dashed blue curve is a guide to the eye showing the peak frequency shift. (c) Peak frequency as a function of the lattice constant (disk-center distance). Three of the data points are from Fig. 2(b) and an additional data point is from a $d$=1.2μm disk array, with frequency scaled to a 0.8μm diameter disk. The solid curve is an exponential fit.

We plot in Fig. 2c the resonance frequencies of Fig. 2b as a function of the disk-disk distance. An additional point from a $d$=1.2 μm disk array, scaled according to the geometrical scaling factor $d^{-1/2}$ of the resonance frequency (see equation (2)), is also included. The red line is a phenomenological exponential decay fitting with a decay constant of 0.45 μm. From the fitting parameter, we find that if the lattice constant (or the disk-center distance) is larger than 2 μm for the 0.8 μm disks (2.5 times the diameter), the coupling of the disks within the same graphene layer is negligible. This is consistent with previous studies of dipole-dipole interactions in metallic nanoparticles [18, 41].

## 3. Vertical plasmonic coupling of graphene micro-disks

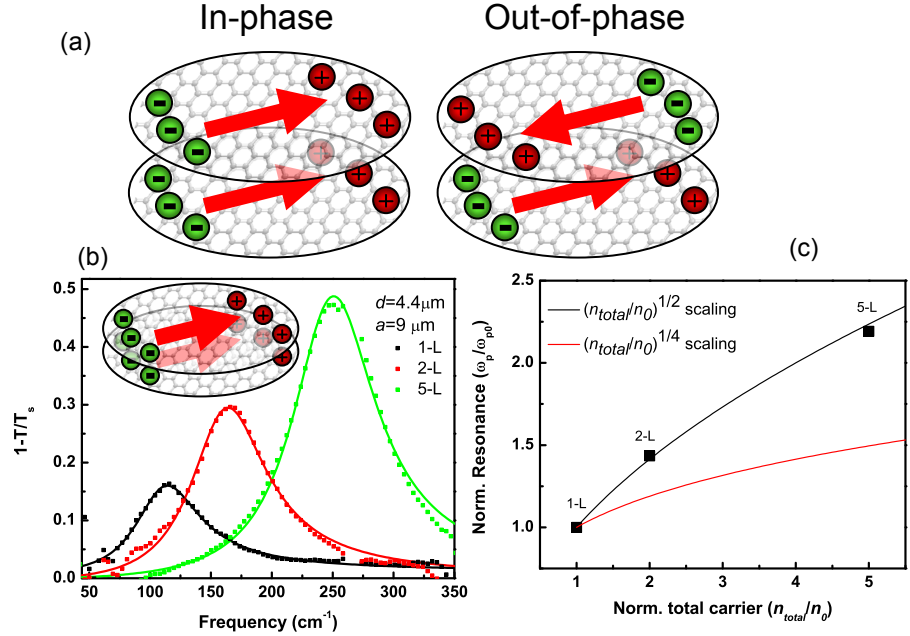

**Figure 3**. Vertical plasmon interaction. (a) Illustration of the in-phase and out-of-phase modes for two stacked disks. (b) Extinction spectra for 1-,2-, and 5-layer graphene disk arrays in the strong vertical coupling limit. The solid curves are fits using Equations (1) and (3). The inset shows a two-layer coupled dipoles. (c) The normalized plasmon frequency as a function of the normalized total carrier density of the 1-,2-, and 5-layer disk arrays whose spectra are shown in (b). The black and red curves are the power law scalings. The $n^{1/2}$ scaling agrees with the results.

The coupling of disks in the same layer is relatively weak. Very strong coupling can be realized for vertically stacked graphene disks. Stacked disks are fabricated through multiple transfers of graphene layers and one final e-beam lithography step. The vertical distance between the stacked disks can, in principle, be controlled by the spacer thickness between the layers. In our case, NFC polymer[43] serves as spacer and many other materials can be used as spacer as well. Fig. 3a depicts the two stacked disks and their collective modes. There are two possible plasmon modes in this two-disk system[44]. The in-phase charge oscillation of the two disks can increase the restoring force and hence the resonance frequency. While on the other hand, the out-of-phase charge oscillation weakens the force and in the limit of zero vertical distance, the restoring force is cancelled. Because the dipoles have opposite sign and the overall dipole is small, the out-of-phase mode does not acquire much oscillator strength.

In our experiments, since the distance between disks (20nm) is much smaller than the disk diameter (microns), we are in the strong coupling limit. The out-of-phase mode doesn't appear in our spectral range due to minimal oscillator strength and low frequency. Fig. 3b shows the spectrum for 1-, 2-, and 5-layer disk arrays on quartz substrates. Except for the layer number, other parameters of the samples such as doping level in each layer, disk diameter and lattice constant are the same. The peak frequency increases dramatically with layer numbers, in conjunction with an enhancement of the resonance amplitude.

Fig. 3c shows the resonance frequency normalized by the single layer value as a function of the total carrier density normalized by the single layer density. It follows a $n^{1/2}$ scaling ($n$ is the carrier density) quite well (black solid curve in Fig. 3c). This is surprising since in a single layer graphene, the plasmon resonance frequency has a $n^{1/4}$ carrier density dependence due to its linear band structure[9, 10]. For comparison, a $n^{1/4}$ scaling curve is also plotted (red curve). Our result indicates that the plasmon resonance frequencies (and amplitudes) from the same number of carriers residing in one layer or in multiple layers are different. Distributing the carriers from one layer into multiple layers can drastically enhance the resonance frequency and amplitude, even though the total carrier density is the same. Since the dependence of plasmon resonance frequency and amplitude on the carrier density for a single graphene layer is relatively weak (frequency $\omega_p \sim n^{1/4}$) compared to that of the conventional two dimensional electron gas (2DEG) systems ($\omega_p \sim n^{1/2}$)[34, 45, 46], increasing the layer number of graphene provides a convenient way to achieve broader tunability.

The $n^{1/2}$ dependence shown in Fig. 3c for the stacked disks can be understood even from a classical point of view. For simplicity, we only consider two closely stacked disks. In this case, since the carrier density is doubled, the restoring force of charge carriers is doubled. Therefore, as an oscillator, the oscillation frequency increases to $\sqrt{2}$ times of the original frequency. However, the carrier density dependence in a single layer can not be understood in this way. For one layer, if the carrier density doubles, the restoring force will double, but the oscillation frequency is not $\sqrt{2}$ times but rather $2^{1/4}$ times of the original frequency[10]. In conventional 2DEG, the two-layer and one layer cases give the same result as long as the two layers are in the strong coupling limit[47, 48]. This difference between graphene and conventional 2DEG originates from the quantum effect of Dirac plasmons, as detailed by Hwang *et al.*[49, 50].

To have a quasi-classical description of plasmons in graphene, we can introduce a carrier density dependent "plasmon mass" $m_p = E_f / v_f^2 \propto n^{1/2}$ (Einstein relation)[51, 52]. This mass can be directly measured through cyclotron resonance[52, 53]. If we distribute carriers from one disk into multiple closely stacked disks, even though the restoring force is still the same, the plasmon mass decreases due to a lower Fermi energy, this will increase the oscillation frequency. In a conventional 2DEG with parabolic band structure, the mass is well-defined and density independent. As a result, the plasmons in conventional 2DEG and 2DEG superlattices[47, 48] can be well described by classical models.

**4. Plasmon hybridization in graphene rings**

After investigating the simple disk geometry, we also studied the plasmon resonance in graphene rings. For the purpose of a stronger signal, we use two-layer closely stacked graphene with the same doping. Fig. 4a shows a SEM image of the sample with graphene rings arranged in a triangular lattice. In the following discussion, we neglect the interaction between individual rings on the same layer.

A plasmon hybridization model, which mimics the concept of molecular orbits, has been developed to explain the plasmon mixing in metal nanostructures with greater complexity than the elementary shapes, such as spheres, disks and spherical voids[14, 25]. It is an intuitive and widely accepted model and frequently used to guide the design and study of plasmonic nanostructures. The plasmons in a graphene ring can be treated as the plasmon hybridization from a graphene disk and a smaller diameter anti-dot. Fig. 4b describes the energy-level diagram for the hybridization. The interaction of the dipole resonances from a disk and an anti-dot in the middle forms two hybrid modes: one is the symmetric (bonding) mode which has in-phase dipole oscillation and lower resonance energy; the other is the anti-symmetric (anti-bonding) mode with out-of-phase dipole oscillation and higher resonance energy. Since the total dipole is larger for the symmetric mode, it has stronger coupling to the electromagnetic field and larger oscillator strength, therefore, it's also called a super-radiant mode. On the other hand, the anti-symmetric mode is sub-radiant with weaker oscillator strength. Indeed, these features are what we observe.

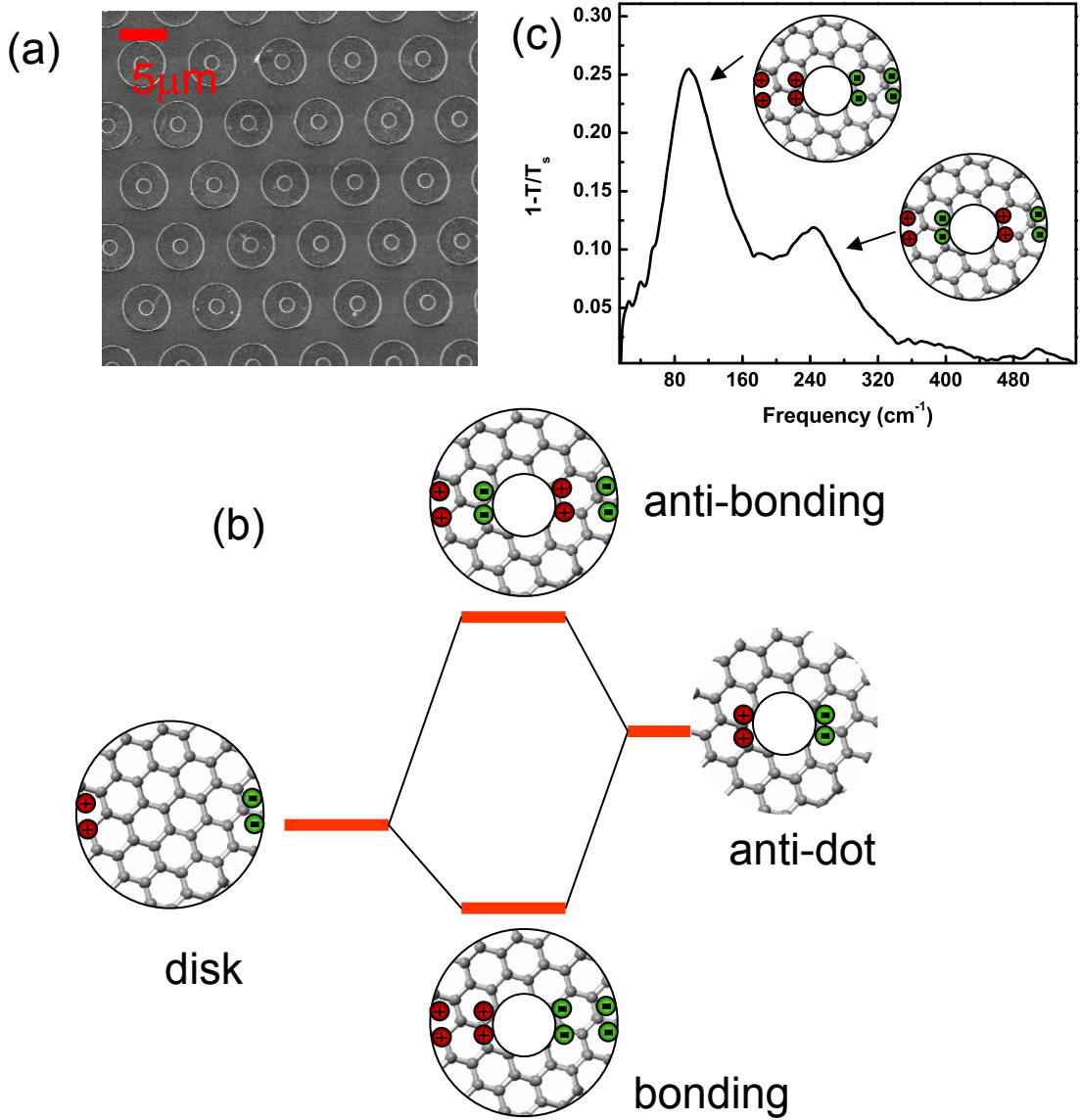

**Figure 4**. Plasmon hybridization in graphene rings. (a) A SEM image of the graphene ring array. The scale bar is 5 μm (b) An illustration of the energy diagram for the plasmon hybridization of a disk and an anti-dot. The plasmons in a ring can be treated as a result of such hybridization. (c) The extinction spectrum for the ring array shown in (a). Insets shows the charge distributions of the two modes.

Fig. 4c shows the spectrum of a ring array with the inner diameter of 1.5μm and outer diameter of 5μm for the rings (A SEM image of the sample is shown in Fig. 4a.). Two well-defined resonances appear in the spectrum and the lower frequency peak has larger amplitude (super-radiant), which is consistent with the feature of a bonding mode. The higher frequency mode is the anti-bonding (sub-radiant) mode. The charge distributions are also schematically shown in the insets. It should be noted that in metal nanostructures (not necessary in rings), the sub-radiant mode usually shows narrower linewidth than that for the super-radiant mode due to the weaker coupling to EM field and hence a longer radiative lifetime of the plasmon[23, 26]. However, we observe similar linewidths for both modes, which indicates that radiative decay has very minor effect on the plasmon lifetime of the micron size graphene structures. This is fully consistent with the fact that the plasmon linewidth in Fig. 4c is almost the same as the Drude scattering width of unpatterned graphene[6, 11], which is the dominant factor to the graphene plasmon linewidth. We note that, the simultaneous observation of the bonding and anti-bonding modes in ring structures of metals is not always an easy task, because the anti-bonding mode usually has higher energy than the interband transition of the parental metal and the mode is severely damped through electron-hole excitations[22].

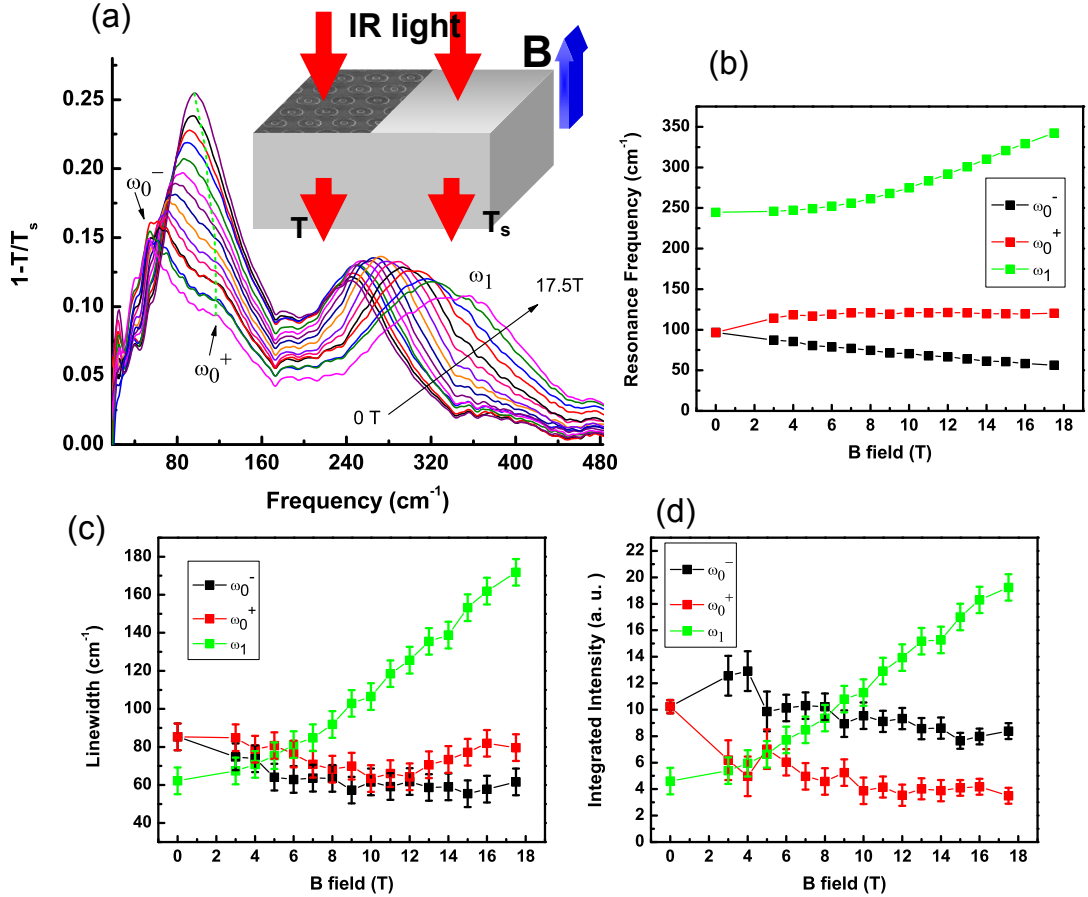

**Figure 5**. Magneto-plasmons in graphene rings. (a) Extinction spectra of the grahene ring array shown in Fig. 4(a) at different magnetic fields from 0 to 17.5 Tesla. Three modes are indicated and the $\omega_0^+$ mode frequency evolution is guided by the green dashed line. The inset depicts the measurement configuration. (b) The frequency dispersions of the three plasmon mode branches. (c) The linewidths of the three modes as a function of the magnetic field. The $\omega_1$ mode broadens significantly. (d) The integrated intensities of the three modes as a function of the magnetic field.

We also studied the plasmon behavior of graphene rings in a perpendicular magnetic field. Unlike plasmons in metals, plasmons in graphene are expected to be strongly affected by a magnetic field due to their small cyclotron mass (usually smaller than $0.1m_0$, $m_0$ is the free electron mass)[52-54] and relatively low plasmon frequencies. In conventional 2DEG systems, the measurement of magneto-plasmons has been extensively employed to explore the effects of many-body interactions[34, 45, 55]. Different geometries of micro- and nano-structures, such as disks[34], rings[56] and elliptical dots[57] were interrogated through the magneto-plasmon excitations. Recently, we studied graphene disks with this technique and a long lifetime of the edge plasmon mode was revealed[16]. Magneto-plasmons were also observed in epitaxial grown graphene on SiC[58].

Fig. 5a shows the extinction spectra for the graphene rings in a magnetic field with strength from 0 to 17.5 Tesla. Salient features include the splitting of the bonding mode into two modes ($\omega_0^-$ and $\omega_0^+$ according to their frequencies), while the anti-bonding mode ($\omega_1$) up-shifts and broadens significantly. We can utilize the previously developed theory for conventional 2DEG rings to qualitative explain the features[59]. The $\omega_1$ branch is a bulk plasmon mode and eventually at the high B-field limit, it becomes the cyclotron resonance of Dirac fermions. Similar to the bulk mode in graphene disks[16], this mode frequency increases (Fig. 5b), its linewidth broadens (Fig. 5c) and gains integrated intensity (Fig. 5d). The $\omega_0^-$ branch is an edge magneto-plasmon mode, similar to the edge mode of a graphene disk. It features a rotating current along the outer circumference. Its frequency gradually decreases with increasing magnetic field strength (Fig. 5b) and the linewidth decreases (Fig. 5c), due to the suppressed back-scattering of the carriers along the edge. Unlike the graphene disk edge mode, however, it gains oscillator strength initially and then loses intensity at high fields (Fig. 5d), an observation consistent with theoretical calculations by Zaremba *et al*.[59]. The $\omega_0^+$ branch features a transition from bulk mode to edge mode. This is manifested in the frequency evolution: first the frequency increases and then levels off, as shown in Fig. 5b. In the high B-field limit, it's an edge mode with a rotating current along the inner circumference. The general trend of the linewidth is decreasing with increasing B-field (Fig. 5c). The oscillator strength decreases from the very beginning, as shown in Fig. 5d. It is worth noting that in a high magnetic field, the plasmon hybridization in the graphene rings is smeared out. This is because the resonances arise from a bulk cyclotron resonance and two independent rotating edge currents with opposite directions localized at the inner and outer circumferences of the rings respectively.

**5. Plasmon coupling to the substrate surface polar phonons**

As a single atomic layer, graphene is very sensitive to the immediate environment, this is especially true for the plasmons in graphene. Excitation of surface polar phonons of the supporting $SiO_2$ substrate is believed to be responsible for the current saturation in graphene devices under certain conditions[60, 61]. In the near field scanning optical microscopy studies of graphene on SiO2, coupled plasmon-surface phonon polariton modes were observed around 1100cm$^{-1}$[12]. Here we show that the plasmons in graphene disks sitting on SiO2 substrate are also coupled to another phonon[62] of SiO2 with the resonance frequency around 500cm$^{-1}$.

Fig. 6a displays the extinction spectra for a graphene disk array on SiO2 substrate at three different doping levels. Different dopings are achieved through the exposure of the sample to the nitric acid vapor[11]. For the as-prepared samples, the Fermi level is usually around -0.3eV (hole doping). After nitric acid treatment, the Fermi level can increase to above -0.5eV. Two peaks are present in each spectrum of Fig. 6a. Obviously, the optical conductivity described by Equation (1) doesn't account for the second peak. It is a plasmon-surface polar phonon hybrid mode. To verify the identity of this second peak, we did the same measurement for graphene disks on non-polar diamond-like carbon surfaces[63] and no such secondary peak was observed. Calculations by Fei *et al.*[12] show that the plasmon coupling to the surface polar phonon can dramatically modify the plasmon dispersion and an anti-crossing of the plasmon mode and the hybrid plasmon-phonon mode can occur. As shown in Fig. 6a, with higher doping, both the peak intensity and the intensity ratio of the hybrid peak to the main peak increase. This intensity ratio is plotted in Fig. 6b as a function of the main peak frequency. Meanwhile, the hybrid peak frequency also slightly increases, but with much weaker doping dependence than that for the main peak. The results indicate that the closer the main plasmon peak approaches the surface phonon frequency, the higher the oscillator strength of the hybrid plasmon-phonon mode. This is consistent with another experiment finding for different size disks. Fig. 6c shows the spectra for two disk arrays with different disk diameters. The intensity of the hybrid peak for the larger disk is much weaker than that of the smaller disk, even though the main peak is much stronger with a frequency further away from the phonon frequency. All these findings are qualitatively consistent with the observation and calculations by Fei *et al*.[12] for another hybrid mode at ~ 1100cm$^{-1}$. Again, our study underscores the importance of graphene-substrate interactions and shows that the substrate plays an integral part in the performance of a graphene device.

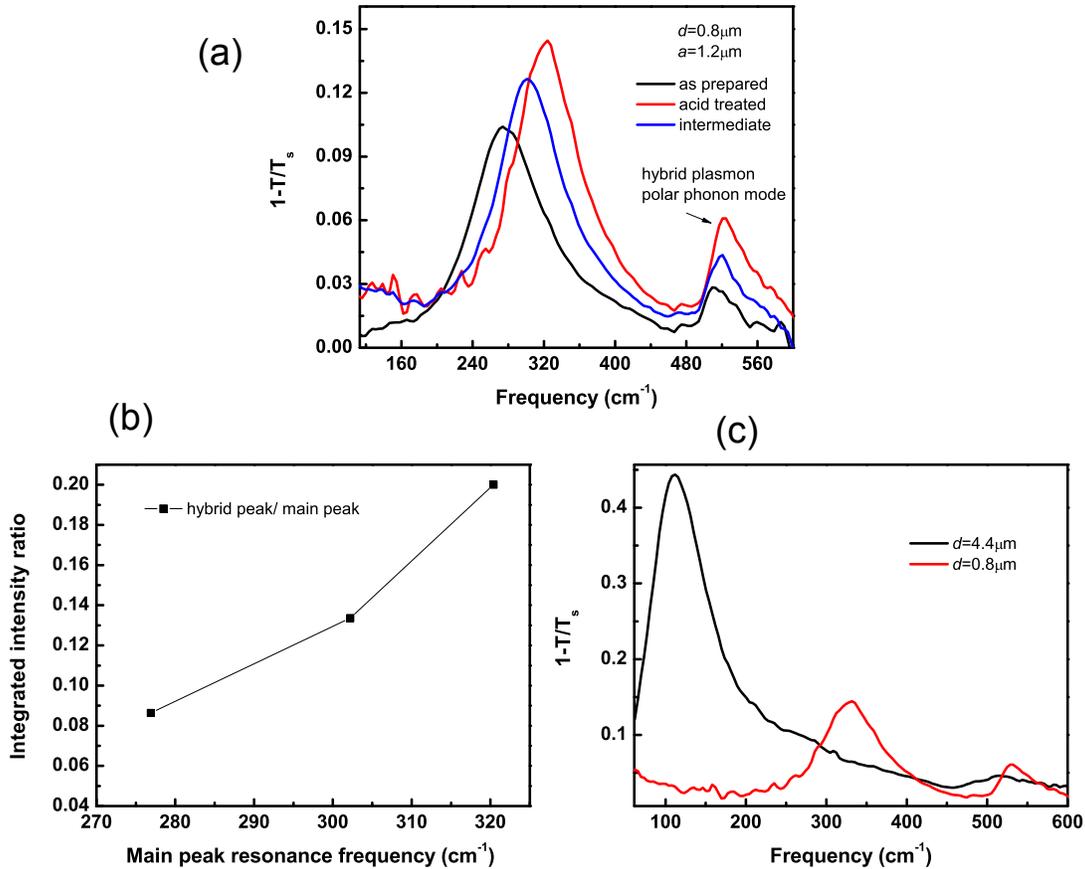

**Figure 6**. The plasmon coupling to the substrate $SiO_2$ surface polar phonons. (a) The extinction spectra of a disk array with $d=0.8\mu m$ and $a=1.2\mu m$ at three different dopings. The peaks around $500 cm^{-1}$ are due to the hybridization of the plasmon mode with a SiO2 surface polar phonon mode. (b) The intensity ratio of the plasmon-phonon hybrid mode to the plasmon mode as a function of the plasmon mode frequency. (c) The extinction spectra for two arrays with different disk diameters. The larger disk has much weaker plasmon-phonon hybrid mode.

## 6. Summary and outlook

To fully explore the potential of graphene as a new plasmonic material, the studies of plasmonic coupling are of great importance. In this paper, four different cases are considered: the coupling of graphene disks on the same plane, the vertical coupling of graphene disks, the plasmon hybridization in graphene rings and the plasmonic coupling to the surface polar phonons. By following the development of the field for plasmons in metals, many more different coupling situations for graphene can be investigated. For instance, it might be feasible to build graphene-based negative index metamaterials[17].

The study of graphene plasmons is not simply a retracing of the development of metal plasmonics. Graphene possesses many unique properties. For instance, as we saw above, the vertical coupling of graphene disks results to unintuitive consequences compared to a single layer disk, due to the density dependent plasmon mass. In the case of coupling to the surface polar phonons, the strength is quite strong due to the one atomic layer thickness of graphene. It is, therefore, foreseeable that graphene may become an important plasmonic material coexisting with other traditional plasmonic materials such as noble metals.

**Acknowledgements**

We thank X. Li, B. Chandra, M. Freitag, G. Tulevski, W. Zhu, Y. Wu, B. Ek and J. Bucchignano for their assistance in the experiments. Part of this work was performed at the National High Magnetic Field Laboratory, which is supported by NSF/DMR-0654118, the State of Florida, and DOE.